\documentclass[a4paper,twocolumn,english,aps,prl,superscriptaddress]{revtex4-1}
\usepackage[T1]{fontenc}
\usepackage[latin9]{inputenc}
\usepackage{color}
\usepackage{babel}
\usepackage{bm}
\usepackage{amsmath}
\usepackage{amssymb}
\usepackage{graphicx}
\usepackage[unicode=true,pdfusetitle,
 bookmarks=true,bookmarksnumbered=false,bookmarksopen=false,
 breaklinks=false,pdfborder={0 0 1},backref=false,colorlinks=false]
 {hyperref}
\usepackage{breakurl}

\makeatletter

\usepackage{babel}
\usepackage{babel}

\makeatother

\begin{document}
\title{Marginal Metals and Kosterlitz-Thouless Type Phase Transition in Disordered
Altermagnets}
\author{Chang-An Li}
\email{changan.li@uni-wuerzburg.de}

\affiliation{Hefei National Laboratory, Hefei, Anhui 230088, China}
\affiliation{Institute for Theoretical Physics and Astrophysics, University of
Würzburg, 97074 Würzburg, Germany}
\author{Bo Fu}
\affiliation{School of Sciences, Great Bay University, Dongguan, 523000, Guangdong
Province, China}
\author{Huaiming Guo}
\affiliation{School of Physics, Beihang University, Beijing 100191, China}
\author{Björn Trauzettel}
\affiliation{Institute for Theoretical Physics and Astrophysics, University of
Würzburg, 97074 Würzburg, Germany}
\author{Song-Bo Zhang}
\email{songbozhang@ustc.edu.cn}

\affiliation{Hefei National Laboratory, Hefei, Anhui 230088, China}
\affiliation{School of Emerging Technology, University of Science and Technology
of China, Hefei, 230026, China}
\date{\today}
\begin{abstract}
Altermagnetism, a recently discovered magnetic phase characterized
by spin-split bands without net magnetization, has emerged as promising
platform for novel physics and potential applications. However, its
stability against disorder---ubiquitous in real materials---remains
poorly understood. Here, we study the electron localization properties
of two-dimensional $d$-wave altermagnets subject to disorder. Remarkably,
we discover a disorder-driven phase transition from a marginal metallic
phase to an insulator, which falls into the Kosterlitz-Thouless class.
We demonstrate this by strong numerical evidence and propose an interpretation
in terms of vortex-antivortex pairs in the disorder-induced local
in-plane spin magnetization. Moreover, we show that the characteristic
spin anisotropy of altermagnets persists but gradually fades away
across the transition. These changes directly affect the spin-splitting
features that are detectable in angle-resolved photoemission spectroscopy
and tunneling magnetoconductance. Our findings provide a new perspective
on recent experimental observations of altermagnetism in candidate
materials.
\end{abstract}
\maketitle
\textit{\textcolor{blue}{Introduction.---}}\textcolor{black}{Altermagnetism
is an emerging collinear magnetic phase} that uniquely combines\textcolor{black}{{}
advantages of ferromagnetism and antiferromagnetism}\ \cite{Smejkal20SACrystal,Naka19NC,Ahn19PRB,Hayami19JPSJ,yuanLD20PRB,Libor22prx1,Libor22prx2,BaiL24AFM,Song25NRM}\textcolor{black}{.
}In momentum space, \textcolor{black}{altermagnets (AMs) exhibit}
highly anisotropic, non-relativistic\textcolor{black}{{} spin splitting}
of electronic bands\textcolor{black}{, e.g., manifesting $d$-wave
symmetric patterns. }From a real-space perspective, this distinctive
feature arises because sublattices with opposite spins are related
by rotational or mirror symmetry, rather than inversion or translation
symmetry\ \cite{Libor22prx1,Libor22prx2}. \textcolor{black}{These
combined properties renders AMs a fertile platform for exploring novel
physics and potential applications in diverse fields such as spintronics,
superconductivity, and topolog}ical phases\ \cite{Rafael21PRL,ShaoDF21NC,Libor22prx3,DiZhu23PRB,ZhangSB24NC,YXLi23PRB,SunC23prb,Papaj23PRB,Beenakker23prb,Ouassou23PRL,Libor23PRL,ZYLiu24PRL,Ghorashi24PRL,ZHouXD24PRL,Leeb24prl,Sato24prl,HPSun25PRB,DuanXK25PRL,GuMQ25PRL,LinHJ25PRL,Antoneko25prl,JXHu25PRL,RChen25arXiv}\textcolor{black}{.
Notably, altermagnetism has been reported in an increasing number
of quantum materials}\ \cite{LeeS24prl,Krempasky24Nature,LiuYC24PRL,ZhuXC25PRL,FYZhang25NP,Ahn19PRB,yuanLD20PRB,Smejkal20SACrystal,Ma2021NC,JiangB25NP,ZhangFY2025RbVTeO,Reimers24NC,DingJY24PRL,ZengM24AS,YangGW25NC,LiuZY24prl,Das24PRL,Xun25APL}.

Despite rapidly growing interest in altermagnetism, the impact of
disorder, which is inevitably present in real materials, on its stability
is barely explored. To date, only a few works have studied the effects
of single impurities in AMs\ \cite{WChen24PRB,Sukhachov24PRB,HRHu25PRB,Maiani25PRB,Gondolf25prb}.
A crucial open question is thus: What is the fate of the AM phase
in presence of disorder? This issue is particularly relevant in light
of recent experiments. For example, in confirmed AM candidates at
high doping levels, such as $\mathrm{Rb_{1-\delta}V_{2}Te_{2}O}$\ \cite{FYZhang25NP},
spin-1/2 dopant atoms may couple \textcolor{black}{opposite electron
spins}, thereby potentially perturbing the underlying AM order. Moreover,
\textcolor{black}{many proposed AM materials, including RuO$_{2}$}\ \textcolor{black}{\cite{Smejkal20SACrystal,Ahn19PRB},
$\mathrm{KV_{2}Se_{2}O}$}\ \textcolor{black}{\cite{Ma2021NC,JiangB25NP,FYZhang25NP}
and $\mathrm{CrSb}$}\ \textcolor{black}{\cite{Reimers24NC,DingJY24PRL,ZengM24AS,YangGW25NC},
are metallic. In such metallic systems, disorder could significantly
alter electronic states and transport properties.}

\begin{figure}
\includegraphics[width=1\linewidth]{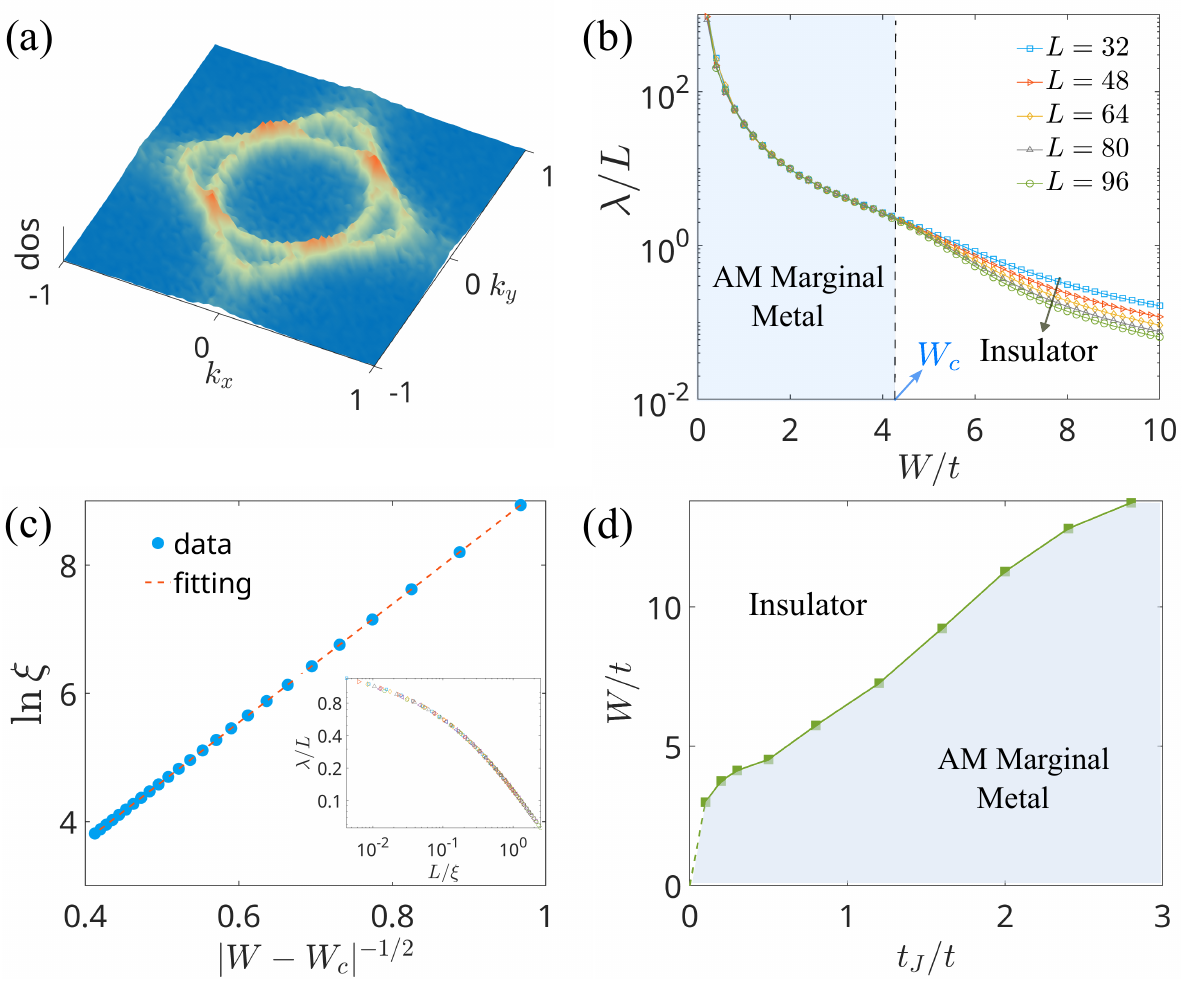}

\caption{(a) Momentum-resolved local density of states (LDOS) of the $d$-wave
AM at disorder strength $W=2t$. (b) Normalized localization length
$\text{\ensuremath{\lambda/L}}$ as a function of $W$ for increasing
system width $L$. $\lambda$ is calculated on a long ribbon with
width $L$ and length of $3\times10^{6}a$. (c) Single-parameter fitting
of the correlation length $\xi$ near the KT-type phase transition,
extracted from the data in (b). On the insulating side, $\xi$ scales
as $\ln\xi\propto|W-W_{c}|^{-1/2}$, where $W_{c}$ denotes the critical
disorder strength. Inset: Collapse of the data from (b) into a single
curve under finite-size scaling. (d) Phase diagram in the $W$-$t_{J}$
plane. Other parameters are $t_{J}=0.3t$ and $\mu=-2t$. \label{fig1:main-result}}
\end{figure}

In this work, we systematically investigate how disorder affects the
stability of the AM phase in two dimensions (2D). Strikingly, we find
that metallic AMs persist as an exotic marginal metal with zero net
magnetization over a finite range of disorder strengths, even when
opposite-spin scattering becomes significant. Beyond a critical disorder
strength, the system undergoes a distinct metal--insulator transition.
Through extensive scaling analyses of correlation length and conductance
in a representative $d$-wave AM system, we find that this phase transition
belongs to the Kosterlitz-Thouless (KT) class. To understand this
transition, we propose a physical picture based on vortex-antivortex
pairs in the induced local in-plane spin magnetization. Furthermore,
the characteristic spin anisotropy persists at weak disorder but gradually
diminishes across the transition. This behavior can be probed via
angle-resolved photoemission spectroscopy (ARPES) and tunneling magnetoconductance.
The results hold for both uncorrelated and correlated disorder and
generalize to $g$-wave AMs and other lattices, for instance, the
Lieb lattice. Our findings underscore the particular impact of disorder
on AMs and suggest disorder as a potential reason for the difficulty
in detecting the predicted spin splitting in candidate materials such
as RuO$_{2}$.

\textit{\textcolor{blue}{Model and general analysis.}}---\textcolor{black}{Altermagnets}
feature alternating spin splitting in electronic bands, related to
spin-space group symmetries\ \cite{Libor22prx2,PFLiu22PRX}. For
concreteness, we consider a disordered $d$-wave \textcolor{black}{AM}
described by the minimal tight-binding Hamiltonian on a square lattice\ \cite{Libor22prx1,Libor22prx2}
\begin{align}
H= & -\sum_{\langle{\bf r},{\bf r}'\rangle,\sigma,\sigma'}\{(t\delta_{\sigma,\sigma'}+t_{J}[{\bf e}_{{\bf r}{\bf r}'}\cdot\bm{\sigma}]_{\sigma\sigma'})c_{{\bf r}\sigma}^{\dagger}c_{{\bf r}'\sigma'}+h.c.\}\nonumber \\
 & -\sum_{{\bf r},\sigma}\mu c_{{\bf r}\sigma}^{\dagger}c_{{\bf r}\sigma}+\sum_{{\bf r},s=0,x,y}c_{{\bf r}\sigma}^{\dagger}w_{{\bf r}}^{s}[\sigma_{s}]_{\sigma\sigma'}c_{{\bf r}\sigma'},\label{eq:Model}
\end{align}
where $c_{{\bf r}\sigma}$ and $c_{{\bf r}\sigma}^{\dagger}$ are
electron annihilation and creation operators at position ${\bf r}$
with spin $\sigma\in\{\uparrow,\downarrow\}$. The notation $\langle{\bf r},{\bf r}'\rangle$
indicates nearest-neighbor sites. The parameter $t$ represents the
normal hopping amplitude and is taken as the energy unit, while $t_{J}$
denotes the spin-dependent alternating hopping strength along different
directions associated with the AM order. The unit vector is ${\bf e}_{ij}=\pm{\bf e}_{z}$
for hopping along $x$ ($y$) direction, $\mu$ is the Fermi energy,
and $\bm{\sigma}=(\sigma_{x,}\sigma_{y},\sigma_{z})$ are the Pauli
matrices for spin.

When disorder is absent, the two spins are decoupled. The system respects
$\text{[}C_{2}||C_{4z}]$ spin symmetry, i.e., a fourfold spatial
rotation about the $z$-axis combined with a spin flip. This symmetry
imposes $d$-wave altermagnets described by $H_{0}({\bf k})=-(2t\cos k_{x}+2t\cos k_{y}+\mu)\sigma_{0}-2t_{J}(\cos k_{x}-\cos k_{y})\sigma_{z}$
with zero net magnetization. The two spin-polarized bands are given
by $E_{\uparrow(\downarrow)}({\bf k})=-2t(\cos k_{x}+\cos k_{y})\mp2t_{J}(\cos k_{x}-\cos k_{y})-\mu$,
where ${\bf k}=(k_{x},k_{y})$ is the 2D momentum and the lattice
constant is set to $a=1$. It gives rise to anisotropically spin-split
Fermi surfaces, as illustrated in Fig.\ \ref{fig1:main-result}(a).

We consider\textcolor{black}{{} onsite disorder} {[}the last term in
Eq.\ \eqref{eq:Model}{]}, which can couple the spin degrees of freedom.
Specifically, the $w_{{\bf r}}^{0}\sigma_{0}$ term represents nonmagnetic
disorder, while $w_{{\bf r}}^{x}\sigma_{x}$ and $w_{{\bf r}}^{y}\sigma_{y}$
correspond to magnetic disorder that flip the spins\ \cite{Libor22prx3}\footnote{Note that the main results we discuss will remain if the disorder
of form $w_{{\bf r}}^{z}\sigma_{z}$ is added. This type of disorder
does not flip the spin}. Here, $w_{{\bf r}}^{s}$ are random values taken from a uniformly
distributed range $[-W/2,W/2]$, with $W$ the disorder strength.
Note that this model explicitly breaks time reversal symmetry, placing
it in the unitary class (class $A$) in 2D\ \cite{Atland97prb,Schnyder08prb}.
Generally, electronic states in this class will be localized even
at infinitesimal disorder strength, except for quantum Hall plateau
transitions\ \cite{WeiHP88prl,Ando83jpsj}. Surprisingly, as we show
below, the AM metals exhibit particular extended states even under
strong disorder, defying this conventional expectation.

\textit{\textcolor{blue}{KT-type transition in disordered AM.}}---To
study the stability of AM metals in presence of disorder, we consider
a long ribbon geometry with width $L_{y}=L$ and length $L_{x}\gg L$.
Whether the states remain delocalized or become localized can be deduced
from the localization length under finite-size scaling. We compute
the localization length $\lambda$ by means of the transfer matrix\ \cite{MacKinnon83ZP,Kramer93RPP,Yamakage13prb},
and define a normalized localization length as $\Lambda\equiv\lambda(L,W)/L$.
According to finite-size scaling theory, $\Lambda$ increases with
$L$ for metallic states, decreases with $L$ for insulating states,
and remains scale-invariant at a critical point\ \cite{MacKinnon83ZP}.

Figure\ \ref{fig1:main-result}(b) shows the scaling behavior of
$\Lambda$ in the disordered AM. Remarkably, over a wide range of
disorder strengths, $\Lambda$ remains independent of $L$, indicating
a marginal metallic state. We refer to this phase as the altermagnetic
marginal metal (AMMM). Its emergence suggests that the AM metallic
state is resilient to disorder up to a considerable strength, which
is crucial for the transport characteristics discussed later. Beyond
a critical disorder strength, however, $\Lambda$ decreases monotonically
as $L$ grows, indicating that the system is turned into an insulator.
It thus realizes a disorder-driven metal-insulator transition. The
appearance of the scale-invariant AMMM phase implies that this transition
belongs to the KT class.

To verify the nature of the transition, we analyze the scaling behavior
of the correlation length $\xi$ (corresponding to $\lambda$ in the
$L\rightarrow\infty$ limit) on the insulating side. In KT-type metal-insulator
transitions, $\xi$ diverges exponentially near the critical disorder
strength $W_{c}$, following a universal form $\xi\propto\exp[b/\sqrt{W-W_{c}}]$,
where $b$ is a constant parameter\ \cite{XieXC98prl,WangC15prl,ChenCZ19prl}.
Indeed, as shown in Fig.\ \ref{fig1:main-result}(c), our numerical
data nicely match the scaling form
\begin{equation}
\ln\xi=b|W-W_{c}|^{-1/2}
\end{equation}
with fitting parameters $b=9.24\pm0.21$ and $W_{c}=4.13t\pm0.08t$.
Moreover, all finite-size scaling curves for different $L$ collapse
to a single curve {[}inset of Fig.\ \ref{fig1:main-result}(c){]}.
These results strongly support that the metal-insulator transition
is of KT type\ \footnote{For KT transition in 2D XY model, the correlation length scales as
$\xi(T)\propto\exp[b/\sqrt{T-T_{c}}]$ above $T_{C}$ with $T$ the
temperature.}. Additionally, by performing the above analysis for different values
of $t_{J}$, we map out the phase diagram in the $t_{J}$-$W$ plane,
shown in Fig.\ \ref{fig1:main-result}(d). The diagram shows that
$W_{c}$ increases with $t_{J}$, indicating enhanced robustness of
the metallic phase by stronger altermagnetism.

\begin{figure}
\includegraphics[width=1\linewidth]{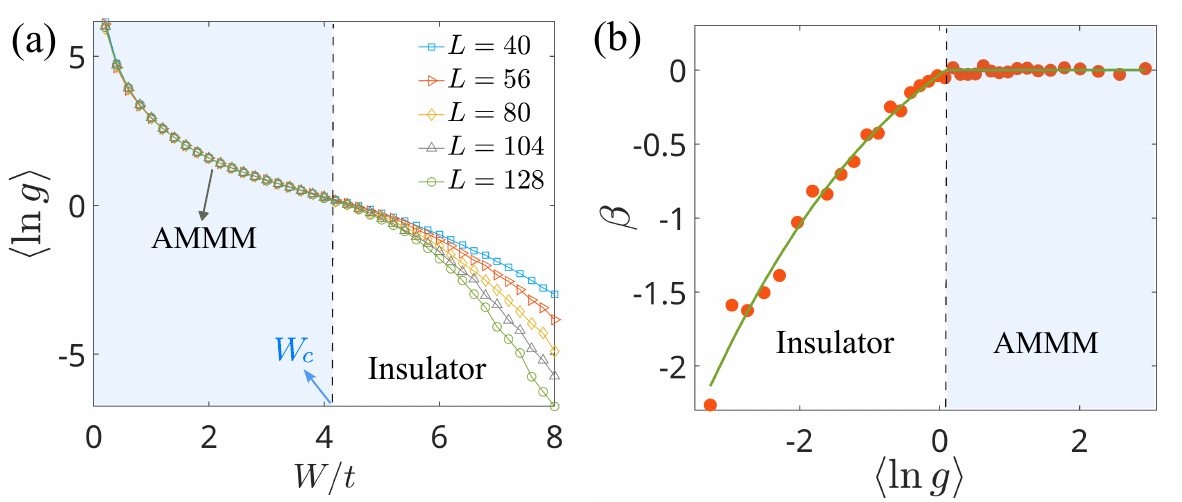}

\caption{(a) Disorder-averaged logarithmic conductance $\langle\ln g\rangle$
as a function of $W$, with each point averaged over 3000 disorder
configurations. (b) Scaling function $\beta\equiv\frac{d\langle\ln g\rangle}{d\ln L}$
extracted from (a). Other parameters are as the same as Fig.\ \ref{fig1:main-result}.}

\label{fig2:conductance}
\end{figure}

\textit{\textcolor{blue}{Scaling features of conductance.}}---Further
evidence for the KT-type phase transition can be obtained from transport
properties. We consider a square AM sample of length $L$, connected
to two conducting leads. The two-terminal conductance $g_{L}$ can
be calculated using the Landauer-Büttiker formalism as $g_{L}=\mathrm{Tr}[TT^{\dagger}]$\ \cite{Dattabook},
where $T$ is the transmission matrix obtained via the recursive Green's
function method\ \cite{MacKinnon85zpbc,Sancho85jpf}. To extract
the intrinsic transport nature, we subtract the contact resistance
between the AM sample and the leads. Thus, the bulk conductance $g$
is defined as $1/g=1/g_{L}-1/N_{C}$, where $N_{C}$ is the number
of conducting channels at the Fermi energy\ \cite{Slevin01prl,ZhangYY09prl}.

Figure\ \ref{fig2:conductance}(a) shows the scaling of the disorder-averaged
conductance $\langle\ln g\rangle$. Remarkably, in the AMMM phase
($W<W_{c}$), curves for different $L$ merge together, indicating
a finite conductance even in the thermodynamic limit. In contrast,
on the insulating side ($W>W_{c}$), $\langle\ln g\rangle$ decreases
monotonically with increasing $L$, implying that all states become
localized. These trends are consistent with the localization length
analysis. To further characterize the transition, we compute the scaling
function $\beta\equiv\frac{d\langle\ln g\rangle}{d\ln L}$\ \cite{Abrahams79prl}.
Typically, a positive (negative) $\beta$ signals delocalized (localized)
states. As shown Fig.\ \ref{fig2:conductance}(b), we find $\beta=0$
in the AMMM region, indicating delocalized states irrespective of
system size. For strong $W(>W_{c})$, $\beta<0$ as expected for the
insulating phase. The analysis of $\beta$ supports the KT metal-insulator
transition in disordered AMs.

\textit{\textcolor{blue}{Spectral statistics.}}---For a comprehensive
understanding of disordered AMs, we examine the spectral statistics
based on level spacing ratio (LSR). The LSR is defined as $r_{n}=\frac{\mathrm{min}(s_{n},s_{n-1})}{\mathrm{max}(s_{n},s_{n-1})}$,
where $s_{n}\equiv E_{n+1}-E_{n}$ is the spacing between adjacent
energy levels $E_{n}$ and $E_{n+1}$\ \cite{Oganesyan07prb}. Figure\ \ref{fig2:statistic}(a)
plots the disorder-averaged LSR $\langle r\rangle$ against disorder
strength $W$. For a finite range of $W<W_{c}$, $\text{\ensuremath{\langle r\rangle}}$
remains nearly constant at $\langle r\rangle\approx0.6$, consistent
with the metallic nature of AMMMs ($\langle r\rangle_{M}=0.6$). As
$W$ exceeds $W_{c}$, $\langle r\rangle$ decreases smoothly toward
$\langle r\rangle_{I}=0.386$, the universal value characteristic
of localized insulating states\ \cite{Atas13prl}. The probability
distributions $P(r)$ at representative $W$ are shown in Fig.\ \ref{fig2:statistic}(b).
For weak disorder ($W=3t$), $P(r)$ closely follows the Gaussian
unitary ensemble (GUE) form $P_{\text{GUE}}(r)=\frac{81\sqrt{3}}{2\pi}\frac{(r+r^{2})^{2}}{(1+r+r^{2})^{4}}$\ \cite{Atas13prl},
which is typical of extended states. For strong disorder ($W=14t$),
$P(r)$ instead matches the Poisson distribution $P_{p}(r)=\frac{2}{(1+r)^{2}}$,
indicative of uncorrelated, localized states. In the SM\ \cite{Li2025SM},
we also analyze inverse participation ratio statistics. These results
further confirm the existence of a robust AMMM phase and the disorder-induced
metal--insulator transition \footnote{Note that there exists a mobility edge in the whole spectrum, but
we focus on the properties at the Fermi energy here}.

\begin{figure}
\includegraphics[width=1\linewidth]{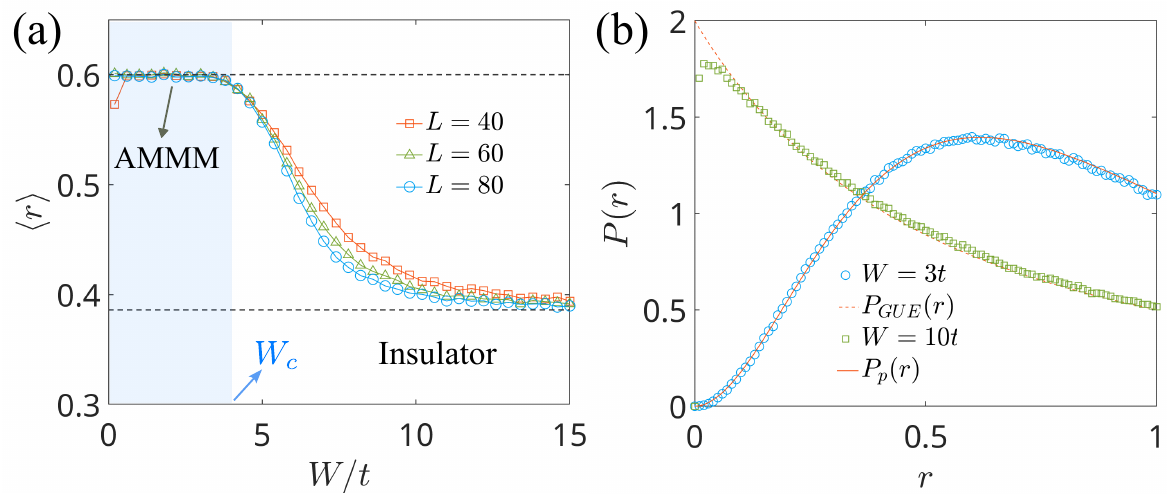}

\caption{(a) Averaged LSR $\langle r\rangle$ as a function of $W$. (b) Distribution
of LSR in the metallic and insulating limits. We take $2\times10^{3}$
disorder configurations. Other parameters are as the same as Fig.\ \ref{fig1:main-result}.}

\label{fig2:statistic}
\end{figure}

\textit{\textcolor{blue}{Phenomenological picture.}}---We present
a phenomenological argument to interpret the KT-type phase transition
in disordered AMs. In the 2D XY model, the KT transition arises from
the unbinding of vortex-antivortex pairs as temperature increases\ \cite{Kosterlitz73JPC}.
Analogously, in 2D AMs, magnetic disorder induces spin flips that
generate pronounced in-plane spin components, whose spatial texture
admits vortices and antivortices\ \cite{Li2025SM}. The disorder
strength $W$ plays the role of temperature in the XY model. At weak
$W$, we can directly visualize the appearance of in-plane spin vortex-antivortex
pairs\ \cite{Li2025SM} corresponding to the marginal metallic phase.
Note that the $[C_{2}||C_{4z}]$ symmetry (equivalently, the $C_{4z}\mathcal{T}$
symmetry that combines four-fold rotation and time-reversal symmetries)
can protect the marginal metallic phase\ \cite{WangF24nc}\footnote{The marginal metal phase can be removed by explicitly breaking $C_{4z}\mathcal{T}$,
for instance, by adding a term $t_{e}(\cos k_{x}+\cos k_{y})\sigma_{z}$
to the Hamiltonian $H_{0}({\bf k})$}. Indeed, as shown in Fig.\ \ref{fig1:main-result}(d), the critical
disorder strength $W_{c}$ increases with the AM strength $t_{J}$.
In absence of AM ($t_{J}=0$) or in conventional antiferromagnets,
no marginal metal phase emerges, underscoring the essential role of
altermagnetism. The marginal metallic phase corresponds to the quasi-ordered
phase in the XY model at low temperatures. When the disorder strength
exceeds $W_{c}$, magnetic scattering between opposite spins becomes
so strong that the in-plane spin vortices proliferate and unpair\ \cite{Li2025SM},
resembling the breakdown of vortex-antivortex pairs in the XY model
at high temperatures. The system then enters an insulating state.
In this way, the AM undergoes a KT-type phase transition driven by
disorder.

\begin{figure}[t]
\includegraphics[width=1\linewidth]{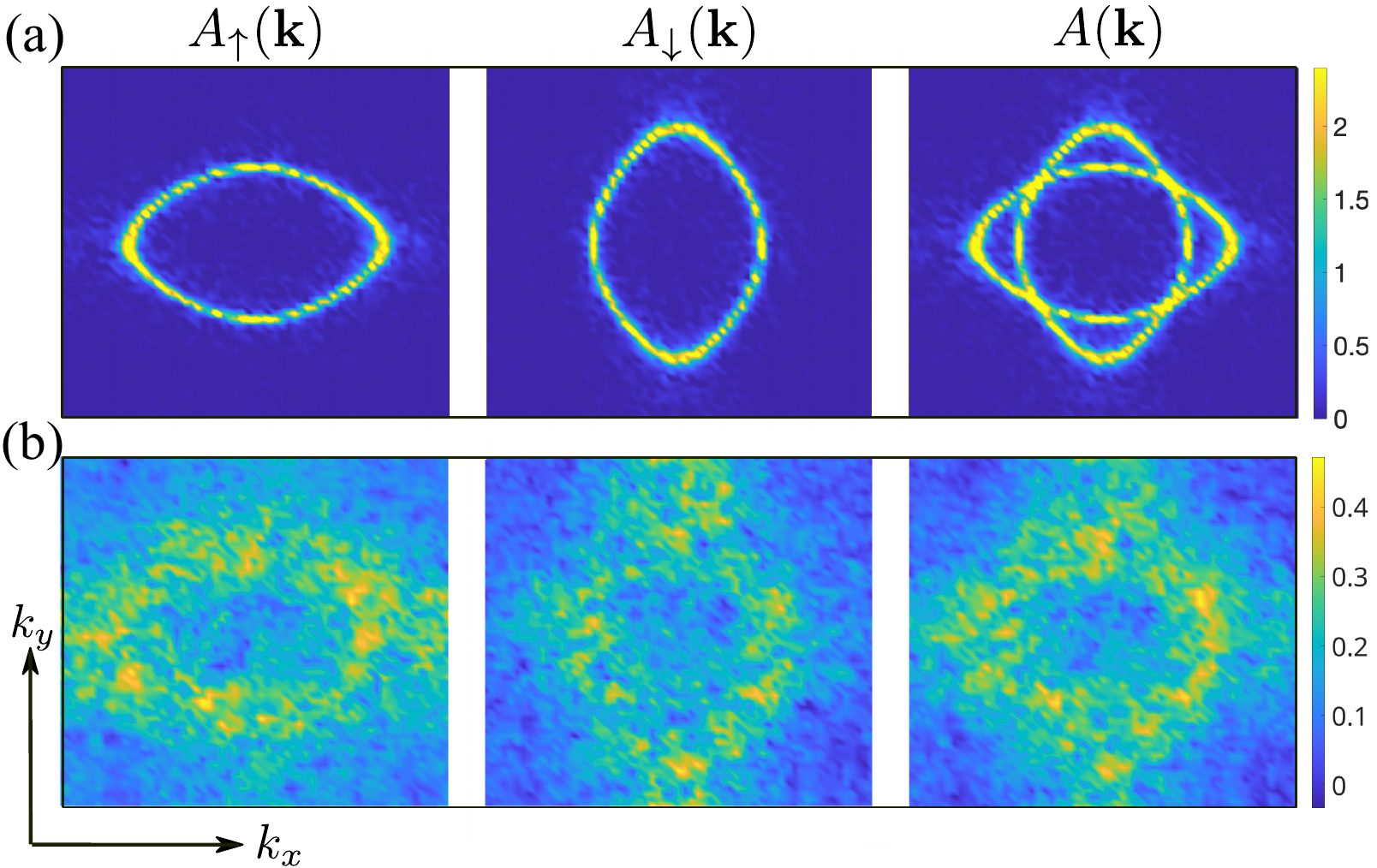}

\caption{(a) Spin-resolved LDOS $A_{\sigma}({\bf k})$ (left and middle) and
total LDOS $A({\bf k})=A_{\uparrow}({\bf k})+A_{\downarrow}({\bf k})$
(right) in momentum space for disorder strength $W=t$. (b) The same
as (a) but for $W=6t$. We take $5\times10^{4}$ disorder configurations.
Other parameters are $t_{J}=0.3t,$ $\mu=-2t$, and $\eta=10^{-3}t$.}

\label{fig3:DOS}
\end{figure}

\textit{\textcolor{blue}{Mitigation of spin anisotropy by disorder.}}\textit{\textcolor{black}{---}}Disorder
significantly affects the characteristic spin anisotropy of AMs, namely,
the alternating spin splitting in momentum space. To illustrate this,
we track the evolution of spin anisotropy during the phase transition
by analyzing spin-resolved LDOS. The LDOS is calculated from the disorder-averaged
effective Hamiltonian $\langle H({\bf k})\rangle$ via
\begin{equation}
A_{\sigma}(E,{\bf k})=-\frac{1}{\pi}\mathrm{Im}[E+i\eta-\langle H({\bf k})\rangle]_{\sigma\sigma}^{-1},
\end{equation}
where $\eta$ is a small spectral broadening\ \cite{Li2025SM}. As
shown in Fig.\ \ref{fig3:DOS}(a), the spin anisotropy remains prominent
in the AMMM phase at weak disorder: spin-up and spin-down Fermi surfaces
exhibit distinct orientations, confirming the persistence of spin-splitting.
As the disorder strength $W$ increases, the bands broaden due to
disorder scattering, and the Fermi-surface splitting gets blurred
{[}see Figs.\ \ref{fig3:DOS}(a,b) for comparison{]}. In this regime,
the anisotropic $d$-wave AM gradually evolves towards an isotropic
pattern. This transition process is similar to the disorder-driven
suppression of $d$-wave superconducting pairing\ \cite{ZhuJX06rmp,Spivak09pcm}.
When $W$ exceeds $W_{c}$, spin anisotropy rapidly fades away {[}Fig.\ \ref{fig3:DOS}(b){]}.
In this strong-disorder regime, spin-up and spin-down states strongly
scatter into each other, and the system eventually enters an insulating
phase. Note that throughout this evolution, the system maintains zero
net magnetization, and spin remains a good quantum number after disorder
average.

The disorder-induced degradation of spin anisotropy in AMs may shed
light on recent experiments on candidate materials such as $\mathrm{RuO_{2}}$
and $\mathrm{MnF_{2}}$\ \cite{Fedchenko24SA,LinZH24arXiv,LiuJY24PRL,Morano25prl}.
In $\mathrm{RuO_{2}}$, while some ARPES measurements report spin-splitting
features\ \cite{Fedchenko24SA,LinZH24arXiv}, others yield conflicting
results\ \cite{LiuJY24PRL}. Previous studies attempt to attribute
this difficulty to potential Ru vacancies\ \cite{Smolyanyuk24PRB}
or insufficient electron interactions\ \cite{QZhuang25PRB}. Our
findings suggest an alternative explanation based on variations in
disorder levels of different samples. In samples with weak disorder,
spin anisotropy is preserved, enabling detection of spin-split bands.
In contrast, in samples with sufficiently strong disorder, spin anisotropy
is fully suppressed, rendering the spin-splitting features unobservable.

\begin{figure}[t]
\includegraphics[width=1\columnwidth]{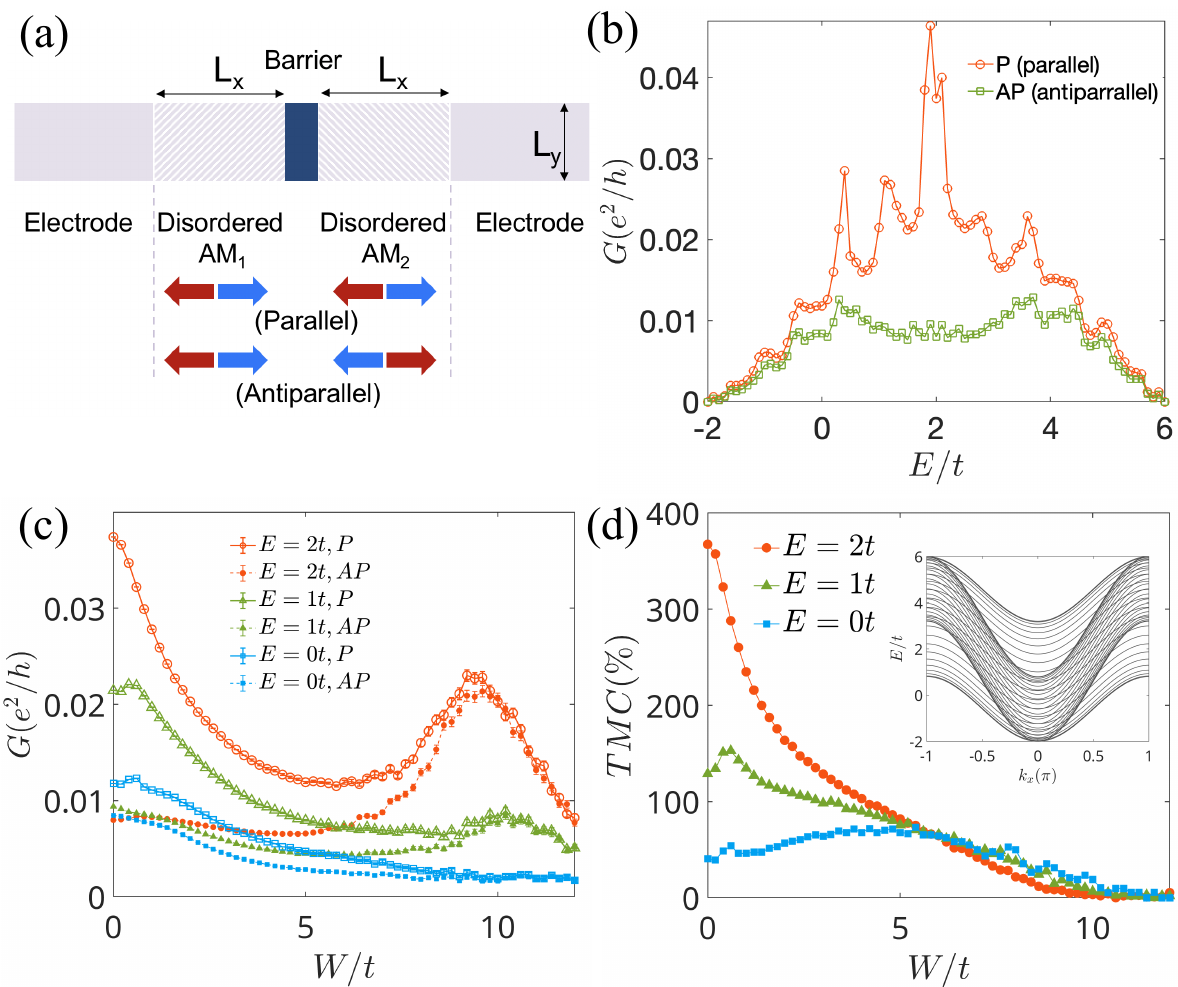}

\caption{(a) Schematic of the tunneling junction composed of two AM layers
of length $L_{x}$ separated by an insulating barrier. The Néel vectors
of two AM layers are configured to be either parallel (P) or antiparallel
(AP). The system size is $L_{x}=10a$ and $L_{y}=40a$. (b) Tunneling
conductances $G_{P}$ and $G_{AP}$ as functions of energy $E$ for
the parallel and antiparallel configurations at $W=0$. (c) $G_{P}$
and $G_{AP}$ as functions of $W$ at different $E$. (d) TMC as a
function of $W$. Inset: Energy spectrum of an AM ribbon of width
$L_{y}=40a$ along $x$ direction. We take $2\times10^{4}$ disorder
configurations. Other parameters are as the same as Fig.\ \ref{fig1:main-result}.
\label{fig:TMR}}
\end{figure}

\textit{\textcolor{blue}{Tunneling magnetoconductance in disordered
AMs.}}\textit{---}The spin splitting affected by disorder in AMs
can also be detected via tunneling magnetoconductance in tunnel junctions.
We consider a junction composed of two AM layers with either parallel
or antiparallel Néel vectors, separated by a tunnel barrier, as illustrated
in Fig.\ \ref{fig:TMR}(a). Recent studies have shown that tunneling
magnetoresistance naturally arises in such junctions owing to the
intrinsic spin splitting \cite{ShaoDF21NC,Libor22prx3,LiuFQ24PRB,NohSH25arXiv}.
Within the energy range of the band width, we calculate the tunneling
conductance using $G=\frac{e^{2}}{h}\mathrm{Tr}[TT^{\dagger}]$, where
$T$ is the transmission matrix across the junction. Due to the spin
anisotropy, the conductance for the parallel configuration $G_{P}$
is generally larger than that of the antiparallel configuration $G_{AP}$,
i.e., $G_{P}>G_{AP}$. This feature is confirmed in Fig.\ \ref{fig:TMR}(b),
which shows $G_{P}$ and $G_{AP}$ as functions of energy. The peaks
in $G_{P}(E)$ originate from tunneling resonances in the junction.

We introduce disorder into the two AM layers and calculate the tunneling
conductances as functions of disorder strength $W$ {[}Fig.\ \ref{fig:TMR}(c){]}.
We find that a substantial conductance difference between $G_{P}$
and $G_{AP}$ persists over a wide range of $W$, extending even to
values $W>W_{c}$. This reflects the fact that spin anisotropy remains
robust against weak disorder in the AMMM phase. However, for strong
disorder ($W\gtrsim10t$), $G_{P}$ and $G_{AP}$ fully converge.
This signals the complete suppression of spin anisotropy due to disorder
scattering. Interestingly, both $G_{P}$ and $G_{AP}$ exhibit nonmonotonic
behavior with increasing $W$, showing an enhancement over a finite
range of $W$ before eventually merging and decreasing. This increase
may be attributed to the fact that AM order reduces while the states
are not fully localized. To quantify the sensitivity of tunneling
to the AM bilayer configuration, we further compute the tunneling
magnetoconductance $\text{TMC}=\frac{G_{P}-G_{AP}}{G_{AP}}$. As shown
in Fig. \ref{fig:TMR}(d), $\text{TMC}$ depends on energy $E$. At
$E=2t$, it decreases monotonically to zero, while at $E=0$, it first
increases to a maximum before it eventually drops to zero. The latter
case suggests that within a specific range of disorder strengths,
the tunneling magnetoconductance can be anomalously enhanced.

\textit{\textcolor{blue}{Conclusion and discussion.---}}In summary,
we have shown that 2D metallic AMs persist as a marginal metal over
a finite range of disorder strengths and undergo a KT metal-insulator
transition as disorder increases further. We have also provided a
phenomenological interpretation of this transition based on spin vortex-antivortex
pairs. During the transition, the spin anisotropy of the AM gradually
weakens, and the tunneling magnetoconductance becomes indistinguishable.
These predictions are experimentally accessible and may help to explain
conflicting experiments on AM candidate materials.

The key ingredient underlying our results is the interplay between
AM order and magnetic disorder. Notably, neither this marginal metallic
phase nor the KT-type metal-insulator transition appears in conventional
antiferromagnets and nonmagnetic electron gases, where the electronic
bands are spin-degenerate. Recently, many 2D AMs have been proposed
and observed experimentally, including Mn$_{2}$PSe with broken inversion
symmetry\ \cite{LiuC25NL}, $\mathrm{Rb_{1-\delta}V_{2}Te_{2}O}$
with layered structures\ \cite{FYZhang25NP,JiangB25NP}, ferroelectric
systems\ \cite{ZhuZ25NL}, monolayer materials\ \cite{MaH21NC}
and twisted magnetic bilayers\ \cite{LiuYC24PRL}. Although our study
focuses on representative $d$-wave AMs, the conclusions hold for
other symmetry types such as $g$-wave AMs and other lattice geometries
such as the Lieb lattice. They remain valid even when the disorder
is spatially correlated\ \cite{Li2025SM}.

We thank Fakher Assaad, Lunhui Hu, K. Mæland, E. Petermann, Jinsong
Xu and Yan-Yang Zhang for helpful discussions. This work was supported
by the DFG (SFB 1170 ToCoTronics), and the Würzburg-Dresden Cluster
of Excellence ct.qmat, EXC 2147 (Project-Id 390858490). C.A.L. was
also supported by the start-up fund at HFNL. H.G. acknowledges support
from the NSFC Grant No. 12074022 and the BNL-CMP open research fund
under Grant No. 2024BNL-CMPKF023. S.B.Z was supported by the start-up
fund at HFNL, the Innovation Program for Quantum Science and Technology
(Grant No. 2021ZD0302801), and the National Natural Science Foundation
of China (Grant No. 12488101).

\textit{Note added}: Recently, we came across a related work that
studies the influence of disorder on $d$-wave altermagnets with strong
spin-orbit coupling and predicts a KT-type phase transition \cite{ChenWW25prb}.
However, our work substantially differs from Ref. \cite{ChenWW25prb}
in that the KT-type transition in our case is solely induced by magnetic
disorder in AMs in the absence of spin-orbit coupling and band topology.

\bibliographystyle{apsrev4-1}

\appendix
\numberwithin{equation}{section}\setcounter{figure}{0}\global\long\def\thefigure{S\arabic{figure}}
\global\long\def\thesection{S\arabic{section}}
\global\long\def\thesubsection{\Alph{subsection}}

\begin{widetext}
\begin{center}
\textbf{\large{}Supplemental materials of ``Marginal Metals and Kosterlitz-Thouless Type Phase Transition in Disordered
Altermagnets''}{\large{} }
\par\end{center}{\large \par}

\section{Localization length and transfer matrix method \label{sec: zLocalization length}}

In this section, we outline the transfer matrix method based on the
lattice model to obtain the localization length of the system\ \citep{MacKinnon83ZP,Kramer93RPP}.
Specifically, we consider a long ribbon geometry along $x$-direction
with width $L_{y}=L$ and length $L_{x}\gg L$. In practice, we take
$L_{x}=3\times10^{6}a$ in units of the lattice constant $a$. Periodic
boundary conditions are taken along $y$-direction. We partition the
ribbon lattice to $L_{x}$ layers along $x$-direction. Focusing on
three consecutive layers at $x=n-1$, $n$, and $n+1$, respectively,
their wave function amplitudes can be connected by

\begin{equation}
\left(\begin{array}{c}
\psi_{n+1}\\
\psi_{n}
\end{array}\right)=T_{n}\left(\begin{array}{c}
\psi_{n+1}\\
\psi_{n}
\end{array}\right),
\end{equation}
 where the transfer matrix is
\begin{equation}
T_{n}=\left(\begin{array}{cc}
-H_{n,n+1}^{-1}(E\bm{I}-H_{n}) & -H_{n,n+1}^{-1}H_{n,n-1}\\
\bm{I} & 0
\end{array}\right).
\end{equation}
Here, $H_{n}$ is the $2L\times2L$ Hamiltonian of the $n$-th layer,
$H_{n,n+1}$ is the $2L\times2L$ interlayer hopping matrix between
the $n$-th and $(n+1)$-th layers, and $\bm{I}$ is a $2L\times2L$
identity matrix.

To extract the localization length, we calculate the transfer matrix
product across the whole ribbon,

\begin{equation}
O_{M}=\prod_{n=1}^{L_{x}}T_{n},
\end{equation}
and construct
\[
P=O_{M}O_{M}^{\dagger}.
\]
According to the Oseledec theorem, positive eigenvalues $\nu_{i}$
of the matrix $P$ exist, and its logarithm defines the Lyapunov exponents
in the limit $L_{x}\rightarrow\infty$ as
\begin{equation}
\gamma_{i}=\lim_{L_{x}\rightarrow\infty}\frac{\ln\nu_{i}}{2L_{x}},
\end{equation}
The quasi-1D localization length is given by the inverse of the smallest
Lyapunov exponent
\begin{equation}
\lambda=\frac{1}{\min(\gamma_{i})}.
\end{equation}

In Fig.~1 of the main text, we show the results for $t_{J}=0.3t$.
Here, in Fig.\ \ref{fig:tJ2.0}(a), we further present the localization
length for $t_{J}=2t$ as a representative case of strong altermagnetic
strength $t_{J}$. Clearly, the critical disorder strength $W_{c}$
shifts to higher values, reflecting the enhanced resilience of altermagnetic
order. By fitting the scaling behavior to the form $\xi\propto\exp[b/\sqrt{W-W_{c}}]$
near $W_{c}$ {[}Fig.\ \ref{fig:tJ2.0}(b){]}, we find $W_{c}=11.27t\pm0.33t$.
All data converge to a single curve under this scaling formula.

\begin{figure}[h]
\includegraphics[width=0.8\linewidth]{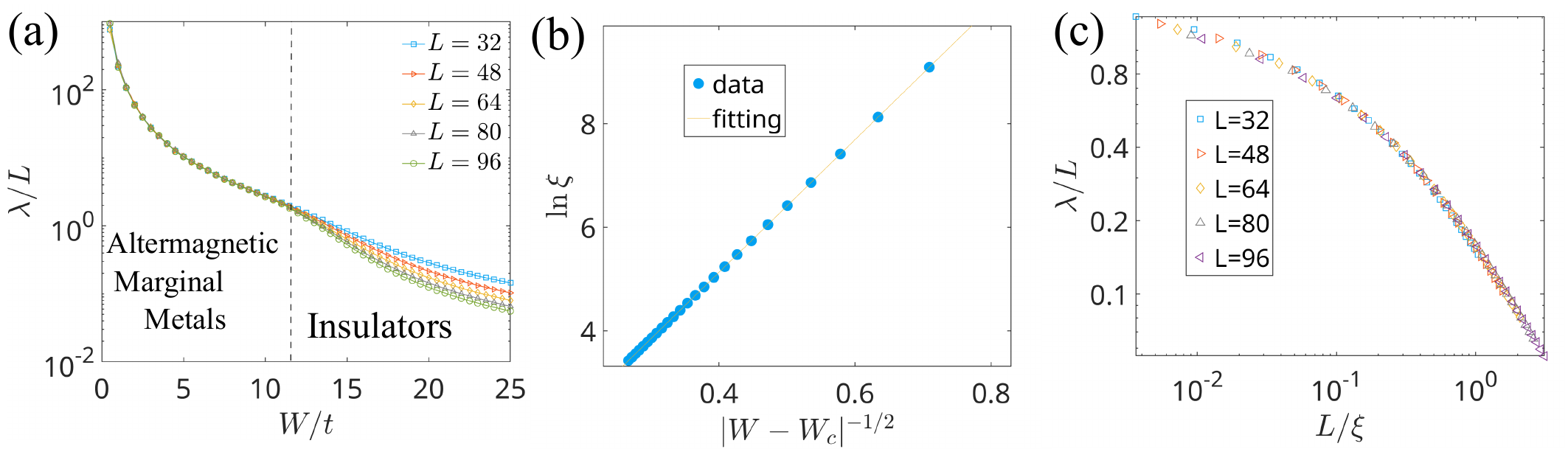}

\caption{Kosterlitz-Thouless (KT) metal-insulator transition for $t_{J}=2t$.
(a) Normalized localization length $\text{\ensuremath{\lambda/L}}$
as a function of $W$ for increasing $L$. (b) Single-parameter fitting
of the correlation length $\xi$ near the KT-type transition, extracted
from the data in (a). It gives the critical disorder strength $W_{c}=11.27t\pm0.33t$.
(c) Collapse of the data from (a) into a single curve under finite-size
scaling. Other parameters are $t_{J}=2t$ and $\mu=-2t$. \label{fig:tJ2.0}}
\end{figure}

\section{Disorder averaged effective Hamiltonian \label{sec: Effective Hamiltonian}}

In this section, we describe in detail how to obtain the disorder-averaged
effective Hamiltonian $\langle H\rangle$. Consider the altermagnetic
metal on a $L\times L$ square lattice with periodic boundary conditions
in both $x$ and $y$ directions. For each disorder configuration,
we find the retarded Green's function $G^{R}({\bf r},{\bf r}';\omega)$
in the tight-binding basis as
\begin{alignat}{1}
G^{R}({\bf r},{\bf r}';\omega) & =\langle{\bf r}|(\omega+i\eta-H_{\text{rand}})^{-1}|{\bf r}'\rangle,
\end{alignat}
where $\omega$ is the frequency, $\eta$ is an infinitesimal positive
number, and $H_{\text{rand}}$ is the lattice Hamiltonian including
disorder. Averaging over many disorder configurations, translation
invariance is effectively restored. Thus, the disorder-averaged Green's
function takes the form
\begin{alignat}{1}
G_{\mathrm{avg}}^{R}({\bf r}-{\bf r}',\omega) & =\langle G^{R}({\bf r},{\bf r}';\omega)\rangle_{\mathrm{dis}}.
\end{alignat}
It depends only on the position difference ${\bf r}-{\bf r}'$. Next,
we Fourier transform the disorder-averaged Green's function into momentum
space,
\begin{alignat}{1}
G^{R}({\bf k},\omega) & =\int d({\bf r}-{\bf r}')G_{\mathrm{avg}}^{R}({\bf r}-{\bf r}',\omega)e^{i{\bf k}\cdot({\bf r}-{\bf r}')}.\label{eq:RetardedG}
\end{alignat}
Finally, the disorder-averaged effective Hamiltonian is then given
by
\begin{equation}
\langle H({\bf k})\rangle=-[G^{R}({\bf k},\omega=0)]^{-1}.
\end{equation}
The spin-resolved density of states follows from
\begin{equation}
A_{\sigma}(E,{\bf k})=-\frac{1}{\pi}\mathrm{Im}[E+i\eta-\langle H({\bf k})\rangle]_{\sigma\sigma}^{-1}.
\end{equation}

\section{Scaling behavior under different disorder types \label{sec: Disorder type}}

In this section, we consider the scaling behavior of disordered altermagnets
(AMs) under different disorder types. In Fig.\ \ref{fig:disorder type}(a),
we consider on-site nonmagnetic (spin-independent) disorder, i.e.,
$w_{{\bf r}}^{0}\sigma_{0}$. Since this type of disorder does not
mix opposite spins, the AMs decouple into two independent spin sectors.
From Fig.\ \ref{fig:disorder type}(a), we find that the normalized
localization length always decreases with system size $L$. This indicates
an immediate transition into the insulating phase and the absence
of the Kosterlitz-Thouless (KT) type phase transition, consistent
with conventional 2D electron behavior. This behavior changes completely
if we consider magnetic disorder of the type $(w_{{\bf r}}^{x}\sigma_{x}+w_{{\bf r}}^{y}\sigma_{y}+w_{{\bf r}}^{z}\sigma_{z})$.
Here, the diagonal type $w_{{\bf r}}^{z}\sigma_{z}$ is included for
comparison, although it does not flip spins. As shown in Fig.\ \ref{fig:disorder type}(b),
$\Lambda=\lambda/L$ remains independent of $L$ within a wide range
of disorder strengths, indicating the existence of a marginal metallic
phase. Beyond the critical value of the disorder strength, $\Lambda$
decreases with $L$, indicating the KT-type metal-insulator transition.
Comparing these results, we conclude that the magnetic scattering
between opposite spins is essential for the occurrence of KT-type
phase transitions in 2D AM metals.

\begin{figure}[h]
\includegraphics[width=0.7\linewidth]{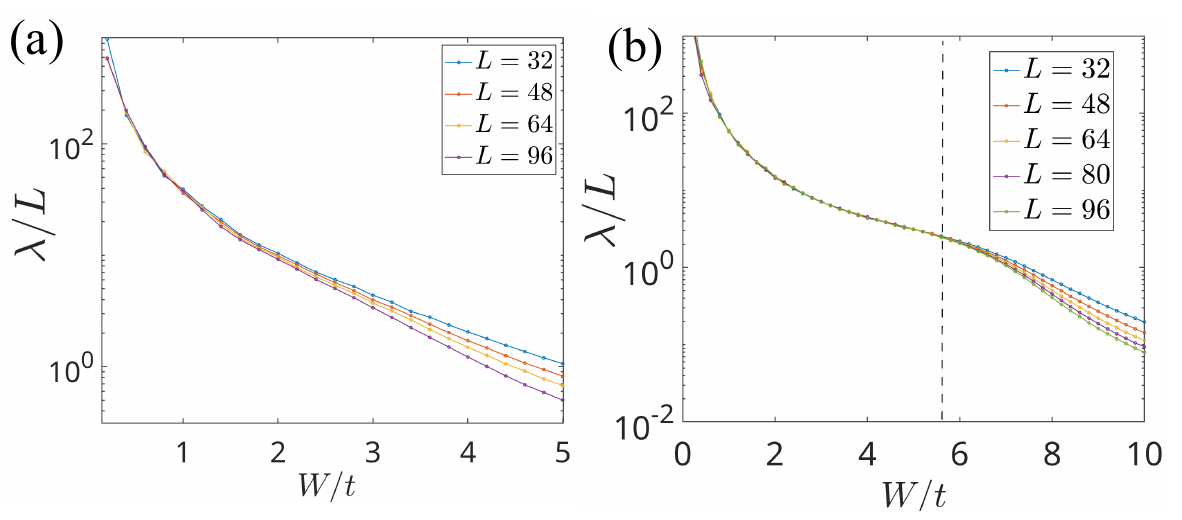}

\caption{(a) Normalized localization length $\text{\ensuremath{\lambda/L}}$
as a function of $W$ for nonmagnetic disorder. (b) Normalized localization
length $\text{\ensuremath{\lambda/L}}$ as a function of $W$ for
magnetic disorder. Other parameters are $t_{J}=0.5t$ and $\mu=-2t$.
\label{fig:disorder type}}
\end{figure}

We further consider correlated disorder of Gaussian type. The disorder
takes the form $V({\bf r})=V_{0}({\bf r})\sigma_{0}+V_{x}({\bf r})\sigma_{x}+V_{y}({\bf r})\sigma_{y}$,
with zero mean $\langle V_{s}({\bf r})\rangle=0$ $(s=0,x,y)$ and
spatial correlations $\langle V_{s}({\bf r})V_{s}({\bf r}')\rangle=(W^{2}/12)\times\exp(-\left|{\bf r}-{\bf r}'\right|^{2}/2\xi^{2})$,
characterized by the correlation length $\xi$ and disorder strength
$W$. A representative disorder configuration is shown in Fig.\ \ref{fig:Gaussian disorder}(a).
To demonstrate the persistence of the marginal metallic phase and
the associated KT-type phase transition, we calculate the scaling
of the conductance as a function of $W$, see Fig.\ \ref{fig:Gaussian disorder}(b).
The conductance curves merge together for different system sizes $L$
in the marginal metallic phase (for $W<W_{c}$) while they decrease
with increasing $L$ in the insulating phase (for $W>W_{c}$) after
a critical disorder strength $W_{c}$.

\begin{figure}[h]
\includegraphics[width=0.7\linewidth]{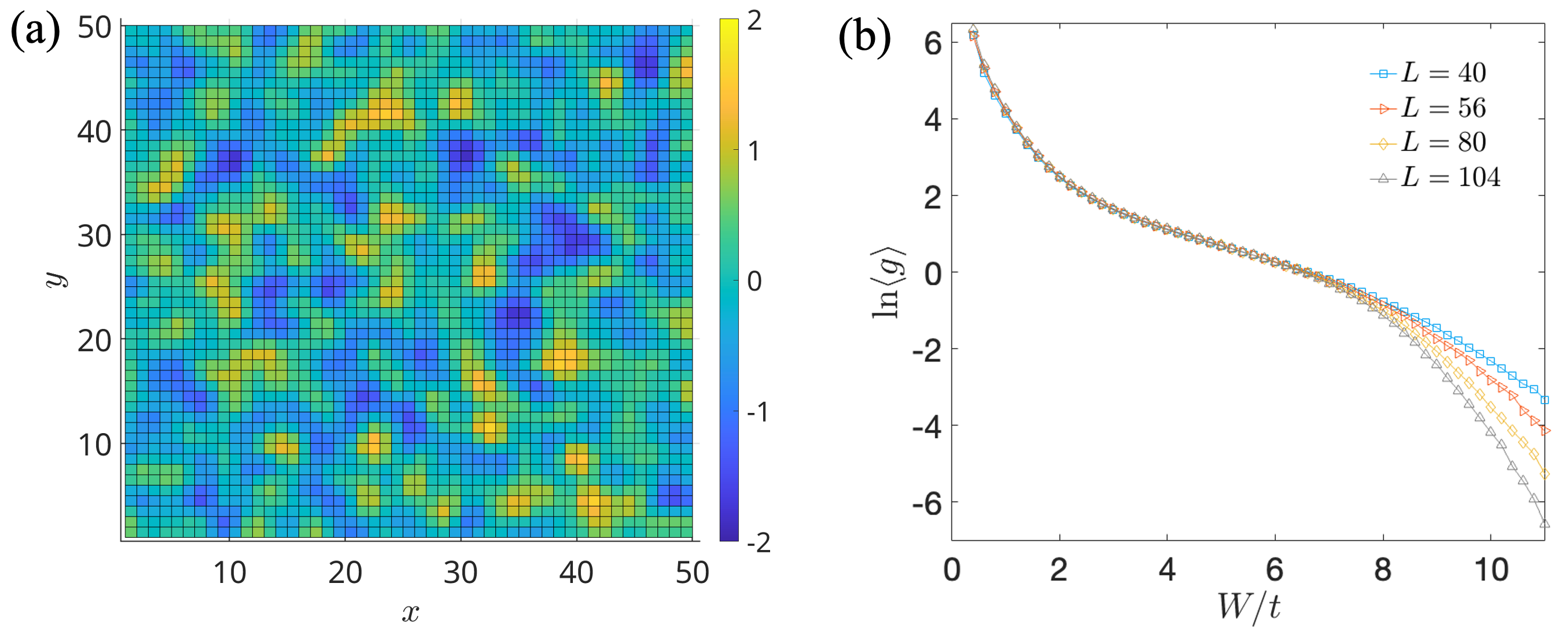}

\caption{(a) A disorder configuration of the Gaussian-correlated disorder in
real space with $\xi=1$ and $W=2t$. (b) Averaged conductance as
a function of disorder strength $W$ for the correlated disorder.
We take $2\times10^{4}$ disorder configurations in the conductance
scaling. Other parameters are $t_{J}=0.3t$ and $\mu=-2t$. \label{fig:Gaussian disorder}}
\end{figure}

\section{Spectrum statistics from inverse participation ratio\label{sec: Statistics}}

In this section, we analyze the statistics of the eigenstates of the
disordered AMs. To this end, we employ the inverse participation ratio
(IPR) defined as\ \citep{LiX17prb,Pixely15prl}

\begin{equation}
I_{n}=\sum_{{\bf r},\sigma}|\psi_{n}({\bf r},\sigma)|^{4},
\end{equation}
where the summation runs over all sites and spins. The disorder-averaged
IPR $\langle I\rangle$ takes nearly zero values at weak disorder
strength, consistent with the altermagnetic marginal metals discussed
in the main text {[}Fig.\ \ref{fig2:statistic}(a){]}. It becomes
finite when the system enters the insulating phase.

The IPR scaling defines a fractal dimension $d_{2}$ via $\langle I\rangle\propto L^{-d_{2}}$.
From Fig.\ \ref{fig2:statistic}(b), we find $d_{2}=1.84$ in the
marginal metal, which is close to the value $d_{2}^{M}=2$ for ideal
metals. For the insulating state at $W=14t$, we find $d_{2}=0.14$,
consistent with the value $d_{2}^{I}=0$ for insulators. Between these
two limits, a smooth transition is observed. Spectral rigidity, quantified
by the spectrum compressibility
\begin{equation}
\chi=\frac{\langle N^{2}\rangle-\langle N\rangle^{2}}{\langle N\rangle}
\end{equation}
also reflects the localization and delocalization behavior. Here,
$\langle N\rangle$ is the disorder-averaged number of energy levels
within an energy window. In 2D, $\chi$ is conjectured to be related
to $d_{2}$ by $\chi=(2-d_{2})/4$. As shown in Fig.\ \ref{fig2:statistic}(c),
we find $\chi=0.035$ at $W=3t$ and $\chi=0.44$ at $W=14t$, both
in excellent agreement with the values inferred from $d_{2}$. Thus,
the conjectured relation is confirmed in both the marginal metallic
and insulating phases.

\begin{figure}
\includegraphics[width=0.8\linewidth]{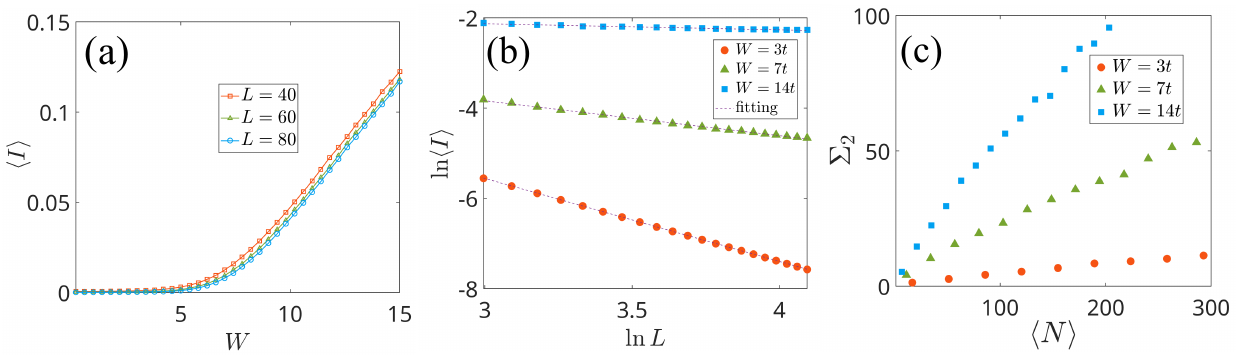}

\caption{(a) Disorder-averaged IPR as a function of disorder strength $W$.
(b) Scaling behavior of IPR for different $W$. (c) Spectrum fluctuation
$\Sigma_{2}\equiv\langle N^{2}\rangle-\langle N\rangle^{2}$ vs $\langle N\rangle$
for different energy windows. Other parameters for all plots are $t_{J}=0.3t$,
and $\mu=-2t$.}

\label{fig2:statistic}
\end{figure}

\section{$d$-wave altermagnets on the Lieb lattice \label{sec: Lieblattice}}

In this section, we consider $d$-wave AMs on a different lattice
geometry, i.e., the Lieb lattice. There are three distinct sites in
each unit cell, labeled as $A,B$, and $C$, respectively, as shown
in Fig.\ \ref{fig5:LiebLattice}(a). The $d$-wave AMs can be described
by the Hamiltonian\ \citep{Brekke23prb}

\begin{equation}
H_{\mathrm{Lieb}}=t\sum_{\langle ij\rangle\sigma}C_{i,\sigma}^{\dagger}C_{j,\sigma}-J\sum_{i,\sigma,\sigma'}C_{i,\sigma}^{\dagger}{\bf S}_{i}\cdot{\bf \bm{\sigma}}_{\sigma\sigma'}C_{i,\sigma'}-\mu\sum_{i,\sigma}C_{i,\sigma}^{\dagger}C_{i,\sigma}+\varepsilon_{B}\sum_{i\in B,\sigma}C_{i,\sigma}^{\dagger}C_{i,\sigma},
\end{equation}
where $t$ is the hopping amplitude, $J$ describes the exchange interaction
strength between itinerant electrons and local spins ${\bf S}_{i}$,
$\mu$ is the chemical potential, and $\varepsilon_{B}$ is the on-site
energy at the nonmagnetic sites. Note that there is no spin-orbit
coupling, thus the Hamiltonian is block diagonal in spin space $\sigma$.
The Bloch Hamiltonian in momentum space for spin $\sigma$ reads

\begin{equation}
H_{\sigma}({\bf k})=\left(\begin{array}{ccc}
-\sigma JS-\mu & t(1+e^{ik_{x}}) & 0\\
t(1+e^{-ik_{x}}) & -\mu+\varepsilon_{B} & t(1+e^{ik_{y}})\\
0 & t(1+e^{-ik_{y}}) & +\sigma JS-\mu
\end{array}\right).
\end{equation}
We plot the energy spectrum along high-symmetry lines in Fig.\ \ref{fig5:LiebLattice}(b).
The energy bands exhibit anisotropic spin splitting, a characteristic
feature of $d$-wave AMs.

Our results of disordered AMs can be generalized to the Lieb lattice.
To illustrate this, we consider the same disorder of the form $(w_{i}^{0}\sigma_{0}+w_{i}^{x}\sigma_{x}+w_{i}^{y}\sigma_{y})$
at each lattice site and perform the scaling of the conductance of
the $d$-wave AM on the Lieb lattice. We find that the conductance
curves for different system sizes $L$ collapse when the disorder
strength $W$ is below a critical value $W_{c}\simeq3t$, signaling
the emergence of a marginal metallic phase. For large $W$ above $W_{c}$,
the conductance decreases with increasing $L$, consistent with an
insulating phase. These results demonstrate the persistence of the
marginal metallic phase and KT-type phase transition in the disordered
$d$-wave AMs on the Lieb lattice.

\begin{figure}
\includegraphics[width=0.8\linewidth]{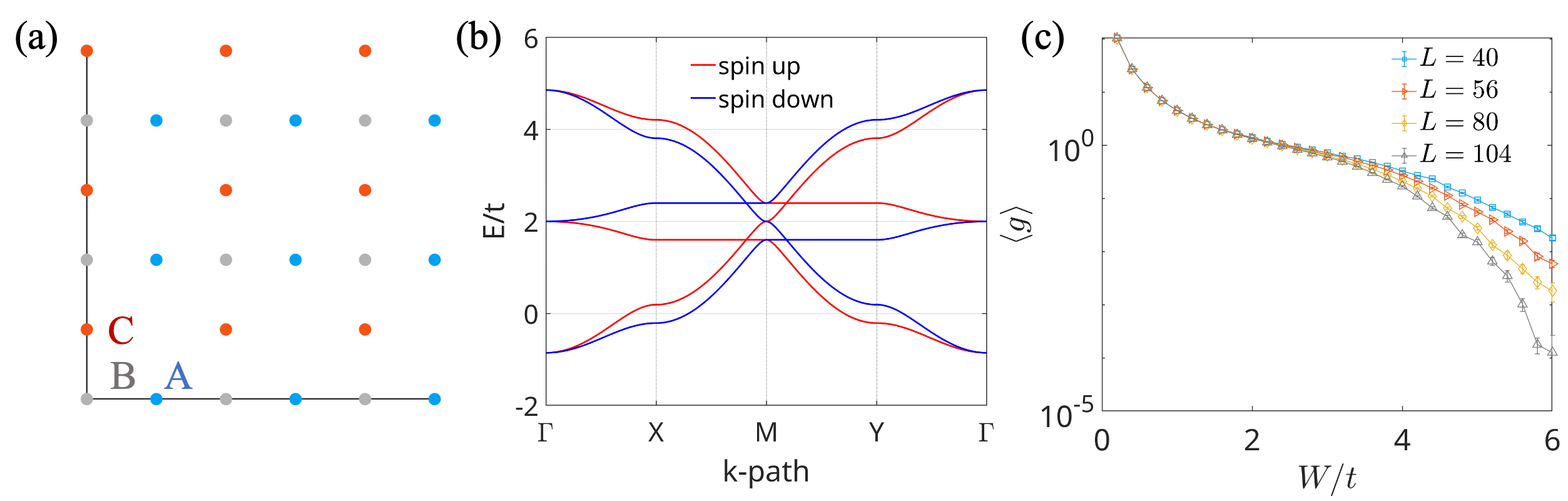}

\caption{(a) Altermagnet in a two-dimensional Lieb lattice, which contains
three distinct sites in each unit cell, labeled by blue, gray, and
red color respectively. (b) The electron bands of AMs on Lieb lattice
along the high symmetry lines. (c) Averaged conductance as a function
of disorder strength $W$. We average over $2\times10^{3}$ disorder
configurations. Other parameters for panels (b) and (c) are $JS=0.6t$,
$\varepsilon_{B}=0$, and $\mu=-2t$.}

\label{fig5:LiebLattice}
\end{figure}

\section{$g$-wave altermagnets and KT-type phase transition\label{sec: g-wave}}

In this section, we show that our results on $d$-wave AMs can be
generalized to $g$-wave AMs in 2D. For illustration, we consider
the minimum effective Hamiltonian for $g$-wave AMs\ \citep{Ezawa25prb}
\begin{equation}
H_{g}({\bf k})=(-\mu+2t-t\cos k_{x}-t\cos k_{y})\sigma_{0}+J\sin k_{x}\sin k_{y}(\cos k_{x}-\cos k_{y})\sigma_{z},
\end{equation}
where $t$ is the usual hopping amplitude, $J$ is the altermagnetic
strength, $\mu$ is chemical potential, and $\sigma_{0}$ and $\sigma_{z}$
are the identity and Pauli matrices in spin space, respectively. The
Fermi surface is shown in Fig.\ \ref{fig:gwave}(a). It clearly exhibits
the g-wave form of spin splitting. We consider the same onsite disorder
of the form $(w_{{\bf r}}^{0}\sigma_{0}+w_{{\bf r}}^{y}\sigma_{y}+w_{{\bf r}}^{z}\sigma_{z})$
as in the main text. We perform the scaling calculation of the conductance
for the $g$-wave AM, as shown in Fig.\ \ref{fig:gwave}(b). The
conductance curves converge for different system sizes $L$ when the
disorder is smaller than a critical values $W_{c}\simeq3.4t$, indicating
the appearance of a marginal metallic phase. While the conductance
decreases as the system size $L$ grows for disorder strength larger
than $W_{c}$. These results demonstrate the presence of the marginal
metallic phase and KT-type phase transition in the disordered $g$-wave
AMs.

\begin{figure}[h]
\includegraphics[width=0.7\linewidth]{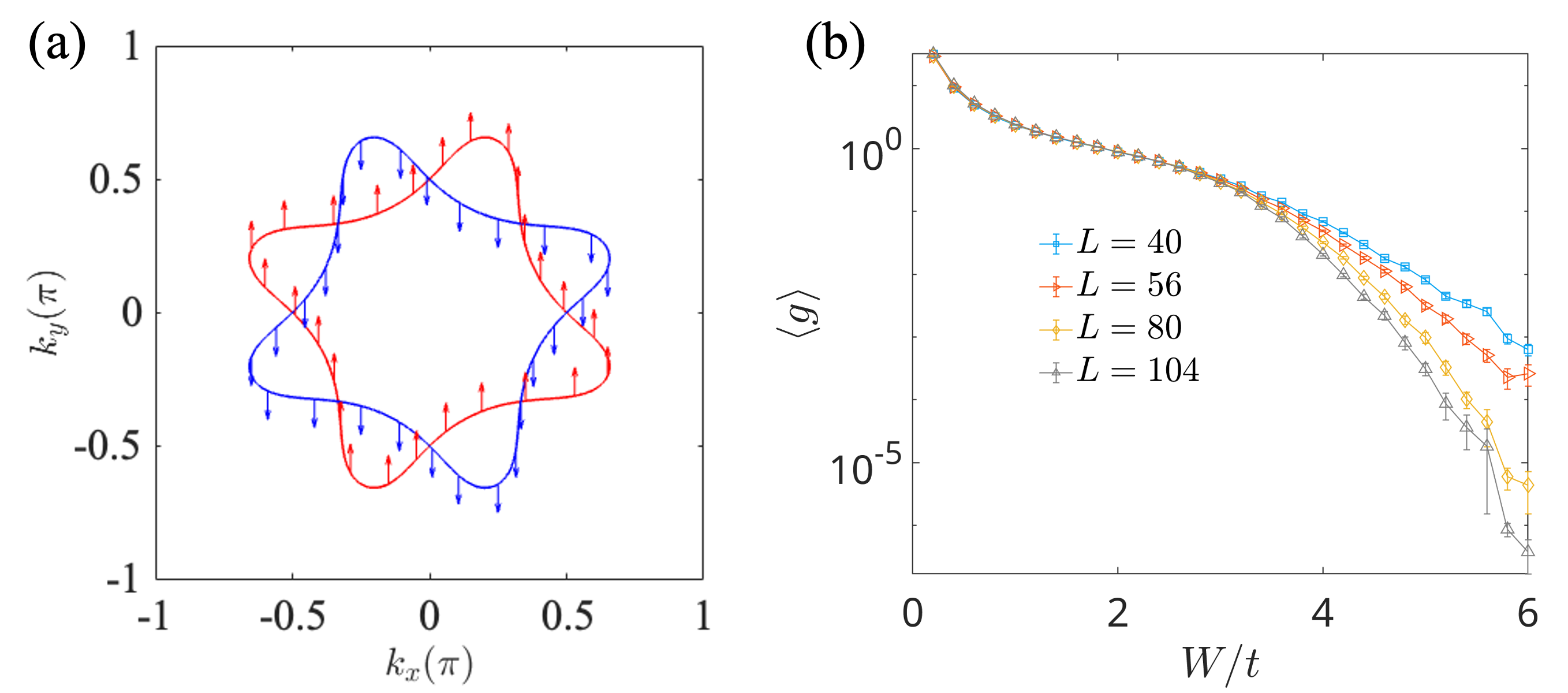}

\caption{(a) Fermi surface of the $g$-wave AM at the Fermi energy, where the
spin-up and spin-down states are indicated by different color and
arrows. (b) Averaged conductance as a function of disorder strength
$W$. We average over $2\times10^{3}$ disorder configurations in
the conductance scaling. Other parameters are $J=1.2t$ and $\mu=-0.6t$.
\label{fig:gwave}}
\end{figure}

\section{Phenomenological picture based on spin vortices \label{sec: Phenomenological Picture}}

In this section, we provide more details to support the phenomenological
picture based on spin vortices in disordered AMs. We first briefly
review the mechanism responsible for the KT phase transition. In the
2D XY model (and thin-film superfluids), the KT phase transition happens
due to the formation of vortex-antivortex pairs\ \citep{Kosterlitz73JPC}.
At zero temperature, there are no vortices in the system. At low temperatures,
bounded vortex-antivortex pairs are thermally activated. In this low-temperature
regime, the system is in a quasi-ordered phase for the XY model and
a superfluid phase for thin-film superfluids. As the temperature increases,
the numbder of vortices increases and vortex-antivortex pairs become
unbounded after a critical temperature, leading to the KT phase transition.
At high temperatures above the transition, the system is in a disordered
phase for the XY model and a normal fluid phase for thin-film superfluid.

The disorder-induced KT-type phase transition in 2D AMs can be explained
along similar lines, with the disorder strength $W$ playing the role
of temperature in the XY model or thin-film superfluid. In the 2D
AMs with strongly anisotropic spin splitting, magnetic disorder of
the form $(w_{{\bf r}}^{0}\sigma_{0}+w_{{\bf r}}^{x}\sigma_{x}+w_{{\bf r}}^{y}\sigma_{y})$
induces spin flips and generate pronounced in-plane magnetization
components whose spatial texture forms a spin texture admitting vortices
and antivortices (phase windings $\text{\ensuremath{\theta({\bf r})}=\ensuremath{\pm}}2\pi$).
The typical in-plane spin textures for a vortex and anitvortex are
schematically shown in Fig.\ \ref{fig5:Phenomenological Picture}(a).
To substantiate this, we calculate numerically the real-space-resolved
in-plane spin components at the Fermi level in presence of magnetic
disorder: 
\begin{equation}
\bar{\sigma}_{s}({\bf r})=\sum_{n}\langle\Psi_{n}({\bf r})\left|\sigma_{s}\right|\Psi_{n}({\bf r})\rangle\rho(E_{n}),
\end{equation}
where $\Psi_{n}({\bf r})$ is the eigenstate, $E_{n}$ the eigenvalue,
and $\rho(E_{n})$ the density of states $\rho(E_{n})=\frac{\Gamma}{E_{n}^{2}+\Gamma^{2}}$
with $\Gamma$ a small band broadening.

We illustrate the results in Fig. \ref{fig5:Phenomenological Picture}(b-d).
At $W=0$, there are no vortices in the system, just as in the 2D
XY model and thin-film superfluid at $T=0$. Remarkably, in the weak-disorder
regime, we observe bound vortex-antivortex pairs appear in the in-plane
spin texture {[}see Fig.\ \ref{fig5:Phenomenological Picture}(b){]}.
As the disorder strength increases, the number of such pairs grows
{[}see Figs.\ \ref{fig5:Phenomenological Picture}(c) and \ref{fig5:Phenomenological Picture}(d){]},
similar to thermally activated vortex pairs in the 2D XY model and
thin-film superfluid. In this case, the system is in a marginal metallic
phase, corresponding to the quasi-ordered phase in XY model or the
superfluid phase in thin-film superfluids at low temperatures. In
the strong-disorder regime, vortices significantly proliferate and
unbind, that is to say, the phase field $\theta({\bf r})$ gets more
and more distorted and in-plane spin correlations become short-ranged
{[}see Fig.\ \ref{fig5:Phenomenological Picture}(d){]}. In this
case, the system is an insulator, corresponding to the disordered
phase in the XY model or the normal fluid phase in thin-film superfluids
at high temperatures. This progression, from bound pairs to vortex
proliferation and vortex unbinding, provides a consistent, microscopic
picture that underscores the KT transition induced by magnetic disorder
in AMs. Note that no vortices appear for purely nonmagnetic disorder
of $w_{{\bf r}}^{0}\sigma_{0}$ type or spin-polarized disorder of
only $w_{{\bf r}}^{x}\sigma_{x}$ type and only $w_{{\bf r}}^{y}\sigma_{y}$
type. Accordingly, no KT-type phase transition occurs under these
types of disorder. This further strengthens our physical picture based
on the formation of in-plane spin vortices.

\begin{figure}
\includegraphics[width=0.7\linewidth]{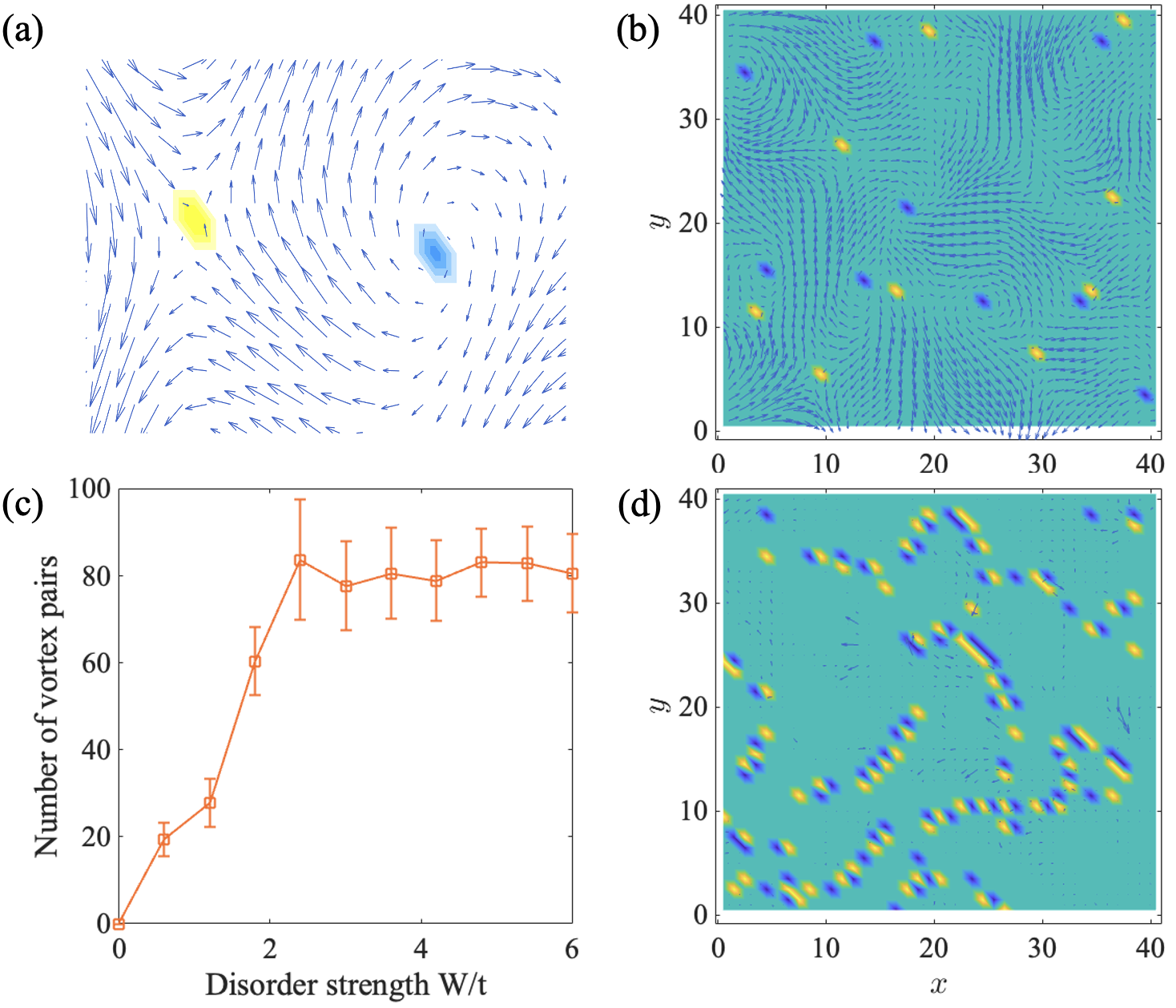}

\caption{Formation of vortex and antivortex in the in-plane magnetization.
(a) Sketch of vortex and antivortex in the in-plane spin components.
(b) Spin texture at delocalized/marginal metal regime with disorder
strength $W=0.5t$. Blue and yellow spots mark vortex and antivortex
cores. (c) Same as (b) but at localized regime with $W=5t$, showing
the proliferation of unbounded vortices and antivortices. (d) Averaged
number of vortex pairs as a function of disorder strength $W$. Here,
50 disorder configurations are taken into the average. We use the
parameters: $t_{J}=0.3t$, $\mu=-2t$, $\xi=2$, and $\Gamma=0.05t$
for illustration.}
\label{fig5:Phenomenological Picture}
\end{figure}

\end{widetext}

\end{document}